\documentclass[showpacs,showkeys,superscriptaddress,aps]{revtex4}
\begin{document}
\title{The nature of manifolds of periodic points for higher dimensional integrable maps}
\author{Satoru SAITO}
\email[email : ]{saito_ru@nifty.com}
\affiliation{Hakusan 4-19-10, Midori-ku, Yokohama 226-0006 Japan}
\author{Noriko SAITOH}
\email[email : ]{nsaitoh@ynu.ac.jp}
\affiliation{Department of Applied Mathematics, Yokohama National University\\
Hodogaya-ku, Yokohama, 240-8501 Japan}
\keywords{Integrable nonintegrable transition, Integrable map, Manifold of periodic points}
\begin{abstract}
By studying periodic points for rational maps on $\bm{C}^d$ with $p$ invariants, we show that they form an invariant variety of dimension $p$ if the periodicity conditions are `fully correlated', and a set of isolated points if the conditions are `uncorrelated'. We present many examples of the invariant varieties in the case of integrable maps. Moreover we prove that an invariant variety and a set of isolated points do not exist in one map simultaneously.
\end{abstract}
\pacs{02.0.+p, 05.45.-a, 45.05.+x, 02.30.Ik}
\maketitle
\section{Introduction}

When a map is given there is no immediate way to foresee the fate of its iterations or whether it is integrable or nonintegrable. If the map is integrable the effect of a small difference of initial values remains small, while it behaves chaotically for some initial values if the map is nonintegrable. In spite of this large difference one cannot easily distinguish the two cases.

It would be desirable to know a way to distinguish integrable maps from nonintegrable ones just by investigation of the first few steps. The purpose of this article is to present a step of finding such a criterion in the case of higher dimensional maps.

In the theory of ordinary 2nd order differential equations, nonexistence of branch point singularities, which depend on initial values, indicates integrability of the equation (the Painlev\'e test). The method of singularity confinement was proposed\cite{GRPH} to replace this test and apply to discrete maps, but it does not always work\cite{HV}. As far as one dimensional maps are concerned there have been intensive studies mainly from the mathematical point of view\cite{Devaney, FM, Beardon}. It is known, for example, that the appearance of a Julia set, the closure of the set of repulsive periodic points, characterizes nonintegrability of the map. On the other hand very little is known about higher dimensional maps. In practice the best we can do at present is to study statistical information of the maps\cite{HV, BV, BS}.\\

We are interested in the transition of a map from a nonintegrable one to integrable one when a parameter in the map is changed continuously. In this paper we pay attention to the behaviour of periodic points of higher dimensional rational maps. If the map is nonintegrable we shall find a set of isolated points with fractal structure as a higher dimensional counterpart of the Julia set. We would like to know what object appears when the map becomes integrable by adjusting the parameter.\\

Before we present the results of this paper let us prepare some words. We assume that the map has $p\ (\ge 0)$ invariants. The periodicity conditions are called `uncorrelated' if they determine all positions of the periodic points dependent on the values of the invariants. They are called `fully correlated' if they do not determine the position of the periodic points but impose relations on the values of the invariants. An `invariant variety of periodic points' is a variety determined by the invariants of the map alone, such that every point on the variety can be an initial point of the iteration, which stays on the variety before it returns to the same point. We would like to show in this paper the following results:

\begin{itemize}
\item Proposition 1

{\it A set of periodic points form a set of isolated points if the periodicity conditions are uncorrelated, and an invariant variety of dimension $p$ if they are fully correlated.}
\item Proposition 2

{\it An invariant variety of periodic points and a set of isolated periodic points do not exist in one map simultaneously.}
\end{itemize}
\vglue 0.5cm

In order to clarify our argument we study in detail the three dimensional Lotka-Volterra (3dLV) map in \S 2 as an example. We will see that the periodicity conditions, which are `uncorrelated' in generic systems, become `fully correlated' in the integrable maps and one gets full varieties of periodic points, instead of isolated points. The proof of our propositions will be given in \S 3. We present in \S 4 more examples of invariant varieties of periodic points of higher dimensional Lotka-Volterra maps and Toda lattice maps. When the dimension $d$ is larger than 5 the invariant varieties become intersections of $d-p$ algebraic varieties. Some examples will be given explicitly in the cases of $d=5$ and 6.

The existence of an invariant variety of some period does not guarantee integrability of the map, but our proposition 2 tells us that there is no set of isolated periodic points. On the other hand, if there exists a set of isolated periodic points of some period, the map can not have an invariant variety but will have infinitely many sets of isolated periodic points, a typical phenomenon of nonintegrable maps. Further discussion about the criterion, however, will be impossible unless we know a precise notion of integrability of a map.\\

Although our results in \S 3 hold in general rational maps, we presented mainly those examples of integrable maps in \S 2 and 4 which belong to the series of Lotka-Volterra maps. The integrability of this series is guaranteed in the sense that all solutions are given explicitly in terms of the $\tau$ function of the KP theory. They are not all of the integrable maps but include a large part of known integrable maps. We explain in the Appendix the relations of the series of Lotka-Volterra maps, the Toda lattice and the $\tau$ function of the KP theory. There is no reason not to study other types of integrable maps.

\section{A study of 3 dimensional Lotka-Volterra map}

Consider a rational map on $\bm{C}^d$
\begin{equation}
\bm{x}=(x_1,x_2,...,x_d)\quad \rightarrow\quad
\bm{X}=(X_1,X_2,...,X_d)=:\bm{X}^{(1)}.
\label{x->X}
\end{equation}
We would like to investigate the behaviour of the sequence: $
\bm{x}\rightarrow \bm{X}^{(1)}\rightarrow\bm{X}^{(2)}\rightarrow\cdots.$ In particular we are interested in the transitions between integrable maps and nonintegrable ones. Many discrete maps have been known to be completely integrable\cite{QRT, Hirota, Suris,  HTI, GR}. Starting from such an integrable map we deform it by introducing a continuous parameter. The map, after the deformation, will be nonintegrable in general. We compare these two systems by changing the parameter and studying their differences.\\

For an illustration let us study the following example\cite{SSYY} with $(x_1,x_2,x_3)=(x,y,z)$ and a parameter $a\in\bm{C}$.
\begin{equation}
X=x{1-y+yz\over 1-z+zx}+a,\quad Y=y{1-z+zx\over 1-x+xy}+2a,\quad Z=z{1-x+xy\over 1-y+yz}.
\label{LV}
\end{equation}
The number of independent terms including $x$ in the numerator of $X$ is four. If we repeat the map it becomes 41 in the numerator of $X^{(2)}$. If we continue repeating it we get 734 of such terms in the numerators of $X^{(3)}$ and more that 20,000 of $X^{(4)}$, {\it etc.}.\\

In order to measure the complexity of the map we can use the algebraic entropy\cite{HV, BV}. Here we define it by
$$
E:=\lim_{n\rightarrow\infty}{1\over n}\ln D_n
$$
with $D_n$ being the highest degree of powers of the initial variable $\bm{x}$ in the numerator after the $n$th map. In the case of (\ref{LV}), $E=\ln 3$. It is known that the map behaves chaotically when $E$ is non-zero in the sense that we fail to predict the long-term behaviour for some initial values.
\\

The situation changes significantly when $a=0$ in (\ref{LV}). In this particular case the map (\ref{LV}) is called the discrete Lotka-Volterra map, and is known to be completely integrable\cite{HTI}. In fact all solutions are given explicitly in terms of a $\tau$ function of the KP hierarchy for arbitrary initial values, as shown in the Appendix.

Let us see what happens in the map (\ref{LV}) when $a=0$. If we repeat the map the number of independent terms including $x$ in the numerator of $X^{(2)}$ is 10. If we continue repeating it we get 68 and 300 such terms in the numerators of $X^{(3)}$ and $X^{(4)}$, instead of 734 and the number of order of $2\times 10^4$ of the case of $a\ne 0$, respectively. This large difference owes to the cancellation of factors in the numerator and the denominator in each step of the map when $a=0$. Correspondingly the highest power of $x$ in the numerator of $X^{(n)}$ changes slowly as 1, 3, 7, 11, 17, $\cdots$, and hence the entropy is zero.

We introduced the parameter $a$ as simply as possible to emphasize that a small deformation from the integrable regime may cause a large difference. Needless to say many other deformations could be chosen. Only if we are very careful do we have a chance to generalize the integrable Lotka-Volterra map without introducing chaos.\\

The algebraic entropy is correlated with an increase of the number of periodic points of the map. In a one dimensional map the Julia set, which characterizes nonintegrability of the map, is defined as the closure of the set of repulsive periodic points. Therefore it will be quite natural to study the behaviour of periodic points to see an indication of the integrability of the map.

If the map returns to the initial point after $n$ steps visiting other points only once, we call the point a point of period $n$. A point of period one is called a fixed point. A periodic point is called attractive (repulsive) if a small neighborhood is mapped closer to (further from) the point. Otherwise the periodic point is called neutral\cite{Devaney}.

To see the problem closer let us first calculate the fixed points of the map (\ref{LV}) when $a\ne 0$. There are two fixed points on the $(x,y)$ plane in $\bm{C}^3$:
\begin{eqnarray}
(x,y,z)&=&\left({2a+3+\sqrt{4a^2+4a+9}\over 4},\ {2a+3-\sqrt{4a^2+4a+9}\over 4},\ 0\right),
\label{fixed p1}\\
(x,y,z)&=&\left({2a+3-\sqrt{4a^2+4a+9}\over 4},\ {2a+3+s\sqrt{4a^2+4a+9}\over 4},\ 0\right)
\label{fixed p2}.
\end{eqnarray}
If we repeat the map we obtain 288 isolated points of period 2.
In general the number of solutions to the set of periodicity conditions
\begin{equation}
X^{(n)}=x,\quad Y^{(n)}=y,\quad Z^{(n)}=z
\label{3 conditions of periods}
\end{equation}
increases very rapidly as $n$ increases. After every step of the map we will get a new set of finite number of `isolated points' in the space of initial values $\bm{x}=(x,y,z)$. From the construction of this set of points, it has a fractal structure as we repeat the map. If we continue this procedure infinitely many times the closure of the set of points might become dense.\\

Now let us turn to the case of an integrable map. In the case of $a=0$ in (\ref{LV}) we find `lines' of fixed points. In fact all points on the four lines
\begin{equation}
\{x=y=0\}\ \cup\ \{y=z=0\}\ \cup\  \{z=x=0\}\ \cup\ \{x=y=z\}
\label{4 cond}
\end{equation}
satisfy the conditions $(X^{(1)},Y^{(1)},Z^{(1)})=(x,y,z)$. This surprises us because we expect only two points at $(x,y,z)=(3/2,0,0),\ (0,3/2,0)$ in the limit of $a\rightarrow 0$ of (\ref{fixed p1}) and (\ref{fixed p2}).
\\

If we repeat the map when $a=0$, we find, instead of 288 points, a hyperbolic surface in $\bm{C}^3$ defined
by
\begin{equation}
\{s+1=0\}.
\label{period 2}
\end{equation}
Here
\begin{equation}
s:=(1-x)(1-y)(1-z).
\label{s}
\end{equation}
is one of two invariants of the 3dLV map\cite{SSYY}. 

This phenomenon continues to the cases of higher period. Namely, if we denote by $r$ the other invariant of the map
\begin{equation}
r:=xyz,
\label{r}
\end{equation}
we find
\begin{equation}
\{r^2+s^2-rs+r+s+1=0\}
\label{period 3}
\end{equation}
for the period 3 case,
\begin{equation}
\{r^3s+s^3-3rs^2+6r^2s+3rs-r^3+s=0\}
\label{period 4}
\end{equation}
for the period 4 case and
\begin{eqnarray}
&&\{r^3s^4-r^3s^2-6r^4s^5+10r^3s^6+3s^5r+s^6+s^5+3r^4s^4-3r^5s^3-6r^4s^3\nonumber\\
&&
-r^6s^3+3r^5s^4+s^4+21s^4r^2+6s^4r+r^3s^7+s^7+27s^5r^2-3s^6r-r^3s^5\nonumber\\
&&
+21r^2s^6-10r^3s^3-6rs^7+s^8=0\},
\label{period 5}
\end{eqnarray}
for the period 5 case, respectively. It is remarkable that the surfaces are characterized only by invariants of the map. These are the invariant varieties of periodic points, since any point on the surfaces can be an initial point and the following iteration of the map remain on the same surface before it comes back to the initial point. It will be possible to find more surfaces associated with higher periods if it is necessary. \\

Now we are going to study the nature of the invariant varieties of periodic points just we have found. The behaviour of a neighbourhood of a periodic point is determined by multipliers $\lambda_1,\lambda_2,$ $...,\lambda_d$, {\it i.e.,} the eigenvalues of the Jacobi matrix of the map\cite{Devaney}. 
\begin{equation}
J=\left(\matrix{{\partial X_1\over \partial x_1}&
{\partial X_1\over \partial x_2}&\cdots&&{\partial X_1\over \partial x_n}\cr
{\partial X_2\over \partial x_1}&{\partial X_2\over \partial x_2}&\cdots&&{\partial X_2\over \partial x_n}\cr
\vdots&&\ddots&&\vdots\cr
{\partial X_n\over \partial x_1}&\cdots&&&{\partial X_n\over \partial x_n}\cr}\right).
\label{J}
\end{equation}
In the directions with $|\lambda_j|<1$ the map is stable while it is unstable in the directions with $|\lambda_j|>1$. If $|\lambda_j|=1$, the point is said to be neutral along the $j$th direction. 

If we calculate eigenvalues for the map (\ref{LV}) with $a\ne 0$ we will see no particular regularity since they are governed by the value of $a$. To study the case of $a=0$ we notice that three equations of the periodicity condition (\ref{3 conditions of periods}) must be satisfied by a single constraint on the $(x,y,z)$ for the periodic points to form an invariant surface. In other words an invariant variety of periodic points of period $n$ emerges when the $n$th image of the map is given as
\begin{equation}
X_j^{(n)}(\bm{x})=x_j+u_j(\bm{x})\gamma_n(\bm{x})\qquad j=1,2,3,
\label{X=x+ugamma}
\end{equation}
with $u_j$'s being regular functions when $\gamma_n=0$. After some manipulation we can convince ourselves that the examples (\ref{period 2}), (\ref{period 3}), (\ref{period 4}) and (\ref{period 5}) of the 3dLV correspond to $\gamma_2,\ \gamma_3,\ \gamma_4$ and $\gamma_5$, respectively.

When (\ref{X=x+ugamma}) holds, the multipliers at a point on the variety $v(\langle\gamma_n\rangle)$ are given by solving
$$
(\lambda-1)^2\left(\lambda-\det J^{(n)}\right)=0,
$$
with
$$
\det J^{(n)}=1+\sum_{j=1}^3u_j{\partial\gamma_{n}\over\partial x_j}.
$$
Therefore it is guaranteed that every point on the variety is neutral along the variety as it should be expected. Moreover $J^{(n)}=1$ holds in the case of the 3dLV. Therefore all points on the invariant surfaces are neutral in all directions. \\

In addition to (\ref{period 3}) there are lines, in the case of period 3, specified by
\begin{equation}
\{x=y=1\}\quad \cup\quad \{y=z=1\} \quad \cup\quad \{z=x=1\},
\label{x=y=1}
\end{equation}
and
\begin{equation}
\{x=y=-\omega\}\quad\cup \quad
\{y=z=-\omega\}\quad\cup \quad
\{z=x=-\omega\}
\label{x=y=-omega}
\end{equation}
\begin{equation}
\cup \quad\{x=y=-\omega^2\}\quad\cup \quad
\{y=z=-\omega^2\}\quad\cup \quad
\{z=x=-\omega^2\},
\label{x=y=-omega^2}
\end{equation}
where $\omega$ is the cubic root of 1. These lines are, however, peculiar in period 3 and not to be considered in the same level of the surface specified by (\ref{period 3}). In (\ref{x=y=1}) the union of the three lines should be conserved. Namely the map does not stay on one line, but goes around following the rule
\begin{equation}
(1,1,t)\rightarrow (t,1,1)\rightarrow (1,t,1)\rightarrow (1,1,t)\rightarrow \cdots
\label{11t}
\end{equation}
with $t\in \bm{C}$. This is due to the discrete symmetry special to the 3dLV map\cite{SSYY}. A point on one of the six lines given by (\ref{x=y=-omega}) and (\ref{x=y=-omega^2}) will not be mapped to the same line either. It corresponds to one of the three singularities at
$$
x(1-y)=1,\qquad y(1-z)=1,\qquad z(1-x)=1
$$
of the map. If we map one of the points, say $(t,1-1/t, z)$, on the surface $x(1-y)=1$ we have a sequence
$$
\left(t,1-{1\over t},z\right)\rightarrow (1,\infty,0)\rightarrow (-\infty,1,0)\rightarrow \left(1-{1\over t},t,z\right)\rightarrow\cdots.
$$
Repeating the map 6 times the point returns to the initial point. It has period 3 if we restrict to $y=x$, {\it i.e.}, when $t=-\omega$ or $-\omega^2$, which are nothing but cases of (\ref{x=y=-omega}) and (\ref{x=y=-omega^2}).\\


\section{The nature of periodicity conditions}

We study in this section the nature of periodicity conditions of a rational map on $\bm{C}^d$ which has $p\ (\ge 0)$ invariants. The periodic points of period $n$ will be found by solving
\begin{equation}
X_j^{(n)}=x_j,\qquad j=1,2,...,d.
\label{periodic conds}
\end{equation}
If $H_1(\bm{x}),H_2(\bm{x}),\cdots,H_p(\bm{x})$ are the invariants, the solutions of (\ref{periodic conds}) are constrained on an algebraic variety of dimension $d-p$ specified by the set of equations
\begin{equation}
H_i(\bm{x})=h_i,\qquad i=1,2,...,p.
\end{equation}
Here $h_1,h_2,...,h_p$ are the values of the invariants determined by an initial point of the map. Let us denote this variety by $V(h)$, {\it i.e.},
$$
V(h)=\Big\{\bm{x}\Big|\ H_i(\bm{x})=h_i,\ i=1,2,...,p\Big\}.
$$

The problem of finding periodic points is equivalent to finding an ideal generated by the set of $d+p$ functions $\{X_j^{(n)}(\bm{x})-x_j,\ \ H_i(\bm{x})-h_i\}$. Since the existence of the invariants enables us to eliminate $p$ components of $\bm{x}$ from (\ref{periodic conds}), the ideal reduces to the $p$th elimination ideal generated by certain functions $\Gamma_n^{(\alpha)}$ satisfying
\begin{equation}
\Gamma_n^{(\alpha)}(h_1,h_2,...,h_p,y_1,y_2,...,y_{d-p})=0,\qquad \alpha=1,2,...,d-p.
\label{Gamma_n=0}
\end{equation}
Here by $y_1,y_2,...,y_{d-p}$ we denote the variables which parameterize the variety $V(h)$ after the elimination of the components of $\bm{x}$.

For an arbitrary set of values of $h_1,h_2,...,h_p$, the functions $\Gamma_n^{(\alpha)}(h,\bm{y})$ define an affine variety, which we denote $V(\langle \Gamma_n\rangle)$, {\it i.e.},
$$
V(\langle\Gamma_n\rangle)=\Big\{\bm{y}\Big|\ \Gamma^{(\alpha)}(h,\bm{y})=0,\ \ \alpha=1,2,...,d-p\Big\}.
$$
In general this variety consists of a finite number of points, hence zero dimension, corresponding to the solutions to the $d-p$ algebraic equations (\ref{Gamma_n=0}) for the $d-p$ variables $y_1,y_2,...,y_{d-p}$. Once the values of $y_1,y_2,...,y_{d-p}$ are decided by solving (\ref{Gamma_n=0}), the location of a periodic point on $\bm{C}^d$ will be determined from the information of the values of $h_1,h_2,...,h_p$. In this way we obtain a number of isolated periodic points of period $n$. This is the precise meaning of the `uncorrelated' periodicity conditions which we introduced in \S 1. Needless to say this case includes a map with no invariant.\\

There are possibilities that the equations $\Gamma_n^{(\alpha)}=0$ impose relations on $h_1,h_2,...,h_p$ instead of fixing all $y_\alpha$'s. If $s$ is the number of independent constraints imposed on $h_i$'s, the same number of $y_\alpha$'s are left free. Accordingly the dimension of the affine variety $V(\langle \Gamma_n\rangle)$ becomes $s$. Every sequence of the map starting from a point on $V(\langle \Gamma_n\rangle)$ will return to the point after $n$ steps.

When $s=d-p$, all $y_\alpha$'s are free, while the values of the invariants are constrained. Let us denote by $V_n(h)$ the variety $V(h)$ whose arguments $(h_1,h_2,...,h_p)$ are constrained by the $d-p$ relations. Then the affine variety $V(\langle \Gamma_n\rangle)$ coincides with the variety $V_n(h)$ itself. In other words every point on $V_n(h)$ is a periodic point of period $n$. Thus all other maps are forbidden on $V_n(h)$, in a strong contrast to generic cases where $s<d-p$ and none of map of other type is excluded on it. In this particular case, we denote by $\gamma_n^{(\alpha)}(h_1,h_2,...,h_p)$ the $d-p$ functions $\Gamma_n^{(\alpha)}(h_1,h_2,...,h_p,y_1,y_2,...,y_{d-p})$ to emphasize independence from the variables $y_1,y_2,...,y_{d-p}$. The functions $\gamma_n^{(\alpha)}$ impose the conditions on $V(h)$, but do not specify a point on it.\\

Let us further replace $h_i$ by $H_i(\bm{x})$ in $\gamma_n^{(\alpha)}$ and write the periodicity conditions as
\begin{equation}
\gamma_n^{(\alpha)}(H_1(\bm{x}),H_2(\bm{x}),...,H_p(\bm{x}))=0,\qquad \alpha=1,2,...,d-p.
\label{gamma=0}
\end{equation}
The expression (\ref{gamma=0}) enables us to consider the constraints on the invariants as constraints on the variables $\bm{x}$. Under this new interpretation of the constraints the values of $h_1,h_2,...,h_p$ are not fixed but only relations among $H_1(\bm{x}),H_2(\bm{x}),...,H_p(\bm{x})$ are imposed to decide the periodic points. If $h_i$'s fulfil the relations $\gamma_n^{(\alpha)}(h_1,h_2,...,h_p)=0$, all points $\bm{x}$ satisfying (\ref{gamma=0}) are periodic points of period $n$. We denote by $v(\langle \gamma_n\rangle)$ the affine variety generated by the functions $\gamma_n^{(\alpha)}(H_1(\bm{x}),H_2(\bm{x}),...,H_p(\bm{x}))$, and distinguish it from $V(\langle \Gamma_n\rangle)$. Namely we define
$$
v(\langle\gamma_n\rangle)=\Big\{\bm{x}\Big|\ \gamma^{(\alpha)}(H_1(\bm{x}),H_2(\bm{x}),...,H_p(\bm{x}))=0,\ \ \alpha=1,2,...,d-p\Big\}.
$$
Note that $v(\langle \gamma_n\rangle)$ is a subvariety of $\bm{C}^d$, whereas $V(\langle \Gamma_n\rangle)$ is a subvariety of $V(h)$, since the values of $h_1,h_2,...,h_p$ are fixed in the latter.

The significance of defining $v(\langle \gamma_n\rangle)$ lies on the fact that for a point to belong to $v(\langle \gamma_n\rangle)$ is a sufficient condition for the point being a periodic point of period $n$. Every point on $v(\langle \gamma_n\rangle)$ can be an initial point of the periodic map of period $n$. This is true only in the case of $s=d-p$. If $s$ is less than $d-p$, the constraints on the invariants are not sufficient to decide periodic points. We can summarize properties of $v(\langle \gamma_n\rangle)$ as follows.
\begin{enumerate}
\item
The dimension of $v(\langle \gamma_n\rangle)$ is $p$.
\item
Every point on $v(\langle \gamma_n\rangle)$ can be an initial point of the periodic map of period $n$.
\item
All images of the periodic map started from a point of $v(\langle \gamma_n\rangle)$ remain on it.
\item
$v(\langle \gamma_n\rangle)$ is determined by the invariants of the map alone.
\end{enumerate}

We have already shown many examples of $v(\langle \gamma_n\rangle)$ in the case of the 3dLV map. We called $v(\langle \gamma_n\rangle)$ in \S 1 an `invariant variety of periodic points'. Since the 3dLV has 2 invariants, the number of independent constraints $s=d-p$ is one, hence the dimension of the invariant varieties is 2. We shall present more examples in \S 4 including the cases of $d-p=2$.

By studying the periodicity conditions (\ref{periodic conds}) we have shown that the nature of periodic map changes largely dependent on the constraints imposed on the invariants. In generic case, where values of the invariants can be chosen arbitrary, the periodic points appear isolated on $\bm{C}^d$. We have called in \S 1 the periodicity conditions being `uncorrelated' in this case. When the invariants are constrained by $d-p$ conditions, on the other hand, the periodic points form an invariant variety of periodic points. This type of conditions was called `fully correlated'. We would say the periodicity conditions being `correlated' if they are not uncorrelated. The results were summarized in Proposition 1 of \S 1.\\

We now suppose that, for a map with $p$ invariants, the periodicity conditions of period $k$ is fully correlated, hence there exists an invariant variety $v(\langle\gamma_k\rangle)$ of periodic points. At the same time we assume that the periodicity conditions of period $n\ (\ne k)$ are uncorrelated, {\it i.e.}, dependent on the variables $y_1,y_2,...,y_{d-p}$. All of the solutions to $\Gamma_n^{(\alpha)}(h_1,h_2,...,h_p,y_1,y_2,...,y_{d-p})=0$ determine points of period $n$ for a set of values of $h_1,h_2,...,h_p$. It must be true even when $h_1,h_2,...,h_p$ are chosen to satisfy $\gamma_k^{(\alpha)}(h_1,h_2,...,h_p)=0$ for all $\alpha$. Recall, however, that all points on $v(\langle\gamma_k\rangle)$ are points of period $k$ irrespective of $y_1,y_2,...,y_{d-p}$. This contradicts to our assumptions that $n\ne k$ and the periodicity conditions of period $n$ are uncorrelated. Therefore the following statements are true for a rational map on $\bm{C}^d$.
\begin{itemize}
\item
{\it If there exists a set of fully correlated periodicity conditions, say of period $k$, all other periodicity conditions of period $n\ne k$ are correlated.}

\item
{\it If there exists a set of uncorrelated periodicity conditions, there exists no set of fully correlated periodicity conditions.}
\end{itemize}

The second proposition of \S 1 follows after these results immediately.

\section{Invariant varieties of higher dimensional maps}

Based on the study of the 3 dimensional Lotka-Volterra map in \S 2, we obtained in \S 3 some results which can be applied to all rational maps in higher dimensions.
Our propositions do not guarantee that an existence of an invariant variety of some period implies integrability of the map. We have not proven either that every integrable map has an invariant variety of periodic points. Nevertheless it will be worthwhile to study many known integrable maps and see if they have invariant varieties of periodic points.

We present in this section some examples of invariant varieties derived from integrable maps whose dimensions are higher than 3. In all examples we shall see that the dimension of the invariant variety coincides with the number $p$ of invariants of the map, in agreement with our result of section 3. The first examples are the Lotka-Volterra map and the Painlev\'e V map whose dimension is 4. The dimension of the invariant varieties is 3 in both cases, while the number of the invariants is also 3. The second example is the 5 dimensional Lotka-Volterra map which has 3 invariants. We shall find an invariant variety of dimension 3 as we expect. The final example is the 3 point Toda lattice map, which is equivalent to the 6 dimensional Lotka-Volterra map. The dimension of the invariant variety is 4, in agreement with the number of the invariants. As the dimension of the map and/or the degree of period increases, the procedure of finding the invariant varieties becomes harder, since we must manipulate algebraic formulae of many variables and/or of higher degree.

\subsection{4 dim Lotka-Volterra map}

The 4 dimensional Lotka-Volterra map is defined by
\begin{eqnarray*}
X_1=x_1{1-x_2-x_3+x_2x_3+x_3x_4\over 1-x_3-x_4+x_3x_4+x_4x_1},&\quad&
X_2=x_2{1-x_3-x_4+x_3x_4+x_4x_1\over 1-x_4-x_1+x_4x_1+x_1x_2},\\
X_3=x_3{1-x_4-x_1+x_4x_1+x_1x_2\over 1-x_1-x_2+x_1x_2+x_2x_3},&\quad&
X_4=x_4{1-x_1-x_2+x_1x_2+x_2x_3\over 1-x_2-x_3+x_2x_3+x_3x_4}
.
\end{eqnarray*}
There are three invariants of the map given by
$$
r=x_1x_2x_3x_4,\qquad t=(1-x_1-x_3)(1-x_2-x_4),
$$$$
u=x_2x_3x_4+x_3x_4x_1+x_4x_1x_2+x_1x_2x_3-x_1x_3-x_2x_4.
$$
We find an invariant variety of periodic points
$$
v(\langle \gamma_2\rangle)=\{\bm{x}|\ t+1=0\}
$$
for the period 2 case, and
$$
v(\langle \gamma_3\rangle)=\{\bm{x}|\ t^3+t^2r-t^2u+t^2-u^2+2rt-u+r+t=0\}
$$
for the period 3 case.

\subsection{Painlev\'e V map}

As an example which does not belong to the Lotka-Volterra series let us examine a discrete analogue of the Painlev\'e V equation given by\cite{Masuda}
\begin{eqnarray}
X_1:=x_1{1-x_2+x_2x_3-x_2x_3x_4\over 1-x_4+x_4x_1-x_4x_1x_2},&\quad &
X_2:=x_2{1-x_3+x_3x_4-x_3x_4x_1\over 1-x_1+x_1x_2-x_1x_2x_3},\nonumber\\
X_3:=x_3{1-x_4+x_4x_1-x_4x_1x_2\over 1-x_2+x_2x_3-x_2x_3x_4},&\quad &
X_4:=x_4{1-x_1+x_1x_2-x_1x_2x_3\over 1-x_3+x_3x_4-x_3x_4x_1}.
\label{Painleve V}
\end{eqnarray}
Using the invariants of this map
\begin{equation}
r=x_1x_2x_3x_4,\quad s=(1-x_1)(1-x_2)(1-x_3)(1-x_4),\quad v=(1-x_2x_4)(1-x_1x_3),
\label{invariants of Painleve}
\end{equation}
we find an invariant variety of periodic points
$$
v(\langle \gamma_2\rangle)=\{\bm{x}|\ s+v=0\}
$$
for the period 2 case and
$$
v(\langle \gamma_3\rangle)=\{\bm{x}|\ (s+v)^2-s(1-r)^2=0\}.
$$
for the period 3 case.

\subsection{5 dim Lotka-Volterra map}

The Lotka-Volterra map $\bm{x}\rightarrow\bm{X}$ of dimension $d$ is defined by solving
\begin{equation}
X_j(1-X_{j-1})=x_j(1-x_{j+1}),\qquad j=1,2,...,d
\label{LV eq}
\end{equation}
for $\bm{X}=(X_1,X_2,...,X_d)$ under the condition $x_{j+d}=x_j$. We show in the Appendix that the number of invariants of this map is
$$
p=[d/2]+1,
$$
where $[d/2]$ is $d/2$ if $d$ is even and $[d/2]=(d-1)/2$ if $d$ is odd. The cases of $d=3,4$, which we studied already, correspond to $p=d-1$. The problem of finding invariant varieties of periodic points becomes more difficult as $d$ increases. \\

By solving (\ref{LV eq}) in the case of $d=5$ we obtain
\begin{eqnarray*}
X_1 &=& x_1{1-x_2-x_3-x_4+x_2x_3+x_2x_4+x_3x_4 +x_4x_5 -x_2x_3x_4-x_2x_4x_5 +x_2x_3x_4x_5
\over 
1-x_3-x_4-x_5+x_3x_4+x_3x_5+x_4x_5+x_5x_1-x_3x_4x_5-x_3x_5x_1+x_3x_4x_5x_1
}\\
X_2 &=& x_2{1-x_3-x_4-x_5+x_3x_4+x_3x_5+x_4x_5+x_5x_1-x_3x_4x_5-x_3x_5x_1+x_3x_4x_5x_1\over
1-x_4-x_5-x_1+x_4x_5+x_4x_1+x_5x_1+x_1x_2-x_4x_5x_1-x_4x_1x_2+x_4x_5x_1x_2}\\
X_3 &=& x_3{1-x_4-x_5-x_1+x_4x_5+x_4x_1+x_5x_1+x_1x_2-x_4x_5x_1-x_4x_1x_2+x_4x_5x_1x_2\over
1-x_5-x_1-x_2+x_5x_1+x_5x_2+x_1x_2+x_2x_3-x_5x_1x_2-x_5x_2x_3+x_5x_1x_2x_3
}\\
X_4 &=& x_4{1-x_5-x_1-x_2+x_5x_1+x_5x_2+x_1x_2+x_2x_3-x_5x_1x_2-x_5x_2x_3+x_5x_1x_2x_3\over 
1-x_1-x_2-x_3+x_1x_2+x_1x_3+x_2x_3+x_3x_4-x_1x_2x_3-x_1x_3x_4+x_1x_2x_3x_4}\\
X_5 &=& x_5{1-x_1-x_2-x_3+x_1x_2+x_1x_3+x_2x_3+x_3x_4-x_1x_2x_3-x_1x_3x_4+x_1x_2x_3x_4
\over 1-x_2-x_3-x_4+x_2x_3+x_2x_4+x_3x_4 +x_4x_5 -x_2x_3x_4-x_2x_4x_5 +x_2x_3x_4x_5}.
\end{eqnarray*}

From our general formula given in the Appendix three invariants of the 5dLV map are
\begin{eqnarray}
H_1&=&x_1x_2+x_2x_3+x_3x_4+x_4x_5+x_5x_1-x_1-x_2-x_3-x_4-x_5,\nonumber\\
H_2&=&x_1x_3+x_2x_4+x_3x_5+x_4x_1+x_5x_2
-x_1x_2x_3-x_2x_3x_4-x_3x_4x_5\nonumber\\
&&-x_4x_5x_1-x_5x_1x_2-x_1x_2x_4-x_1x_3x_4-x_1x_3x_5-x_2x_3x_5-x_2x_4x_5\nonumber\\
&&+x_2x_3x_4x_5+x_3x_4x_5x_1+x_4x_5x_1x_2+x_5x_1x_2x_3+x_1x_2x_3x_4,
\label{H_2 of d=5}\\
r&=&x_1x_2x_3x_4x_5.
\nonumber
\end{eqnarray}
If we form two particular combinations $H_2+3H_1+5$ and $r+H_1+2$ from these invariants, we see that the Gr\"obner basis of these functions generates the 3rd elimination ideal of the functions $\{X_j^{(2)}-x_j\}$. Therefore the invariant variety of periodic points of period 2 is
$$
v(\langle \gamma_2\rangle)=\{\bm{x}|\ H_2+3H_1+5=0,\ r+H_1+2=0\}.
$$
This is an algebraic variety of dimension 3.

\subsection{3 point Toda lattice map}

As we show in the Appendix, the $N$ point Toda map is equivalent to the $2N$ dimensional Lotka-Volterra map. Since it is much easier to manage the Toda map than the corresponding Lotka-Volterra map from the computational point of view, we consider here the 3 point Toda map,
$$
(I_1^{(t)},I_2^{(t)},I_3^{(t)},V_1^{(t)},V_2^{(t)},V_3^{(t)})\ \ \rightarrow\ \ (I_1^{(t+1)},I_2^{(t+1)},I_3^{(t+1)},V_1^{(t+1)},V_2^{(t+1)},V_3^{(t+1)}),
$$
which is defined by
$$
I_1^{(t+1)} = I_2^{(t)}{V_3^{(t)}V_1^{(t)}+I_3^{(t)}I_1^{(t)}+I_3^{(t)}V_1^{(t)}\over I_2^{(t)}V_3^{(t)}+V_2^{(t)}V_3^{(t)}+I_2^{(t)}I_3^{(t)}},\quad
V_1^{(t+1)} = V_1^{(t)}{I_2^{(t)}V_3^{(t)}+V_2^{(t)}V_3^{(t)}+I_2^{(t)}I_3^{(t)}\over V_3^{(t)}V_1^{(t)}+I_3^{(t)}I_1^{(t)}+I_3^{(t)}V_1^{(t)}},
$$$$
I_2^{(t+1)} = I_3^{(t)}{I_1^{(t)}V_2^{(t)}+I_1^{(t)}I_2^{(t)}+V_1^{(t)}V_2^{(t)}\over V_3^{(t)}V_1^{(t)}+I_3^{(t)}I_1^{(t)}+I_3^{(t)}V_1^{(t)}},\quad
V_2^{(t+1)} = V_2^{(t)}{V_3^{(t)}V_1^{(t)}+I_3^{(t)}I_1^{(t)}+I_3^{(t)}V_1^{(t)}\over I_1^{(t)}V_2^{(t)}+I_1^{(t)}I_2^{(t)}+V_1^{(t)}V_2^{(t)}},
$$$$
I_3^{(t+1)} = I_1^{(t)}{I_2^{(t)}V_3^{(t)}+V_2^{(t)}V_3^{(t)}+I_2^{(t)}I_3^{(t)}\over I_1^{(t)}V_2^{(t)}+I_1^{(t)}I_2^{(t)}+V_1^{(t)}V_2^{(t)}},
\quad
V_3^{(t+1)} = V_3^{(t)}{I_1^{(t)}V_2^{(t)}+I_1^{(t)}I_2^{(t)}+V_1^{(t)}V_2^{(t)}\over I_2^{(t)}V_3^{(t)}+V_2^{(t)}V_3^{(t)}+I_2^{(t)}I_3^{(t)}}.
$$

This map has four invariants,
\begin{eqnarray*}
T_1&=&I_1^{(t)}+I_2^{(t)}+I_3^{(t)}+V_1^{(t)}+V_2^{(t)}+V_3^{(t)},\\
T_2&=&I_1^{(t)}I_2^{(t)}+I_2^{(t)}I_3^{(t)}+I_3^{(t)}I_1^{(t)}+V_1^{(t)}V_2^{(t)}+V_2^{(t)}V_3^{(t)}+V_3^{(t)}V_1^{(t)}\\
&&\qquad\qquad\qquad\qquad+I_1^{(t)}V_2^{(t)}+I_2^{(t)}V_3^{(t)}+I_3^{(t)}V_1^{(t)},\\
T_3&=&I_1^{(t)}I_2^{(t)}I_3^{(t)},\qquad T'_3\ =\ V_1^{(t)}V_2^{(t)}V_3^{(t)}.
\end{eqnarray*}

From these data we find the Gr\"obner bases generating the 4th elimination ideal of $\{X_j^{(2)}-x_j\}$ and $\{X_j^{(3)}-x_j\}$. The invariant varieties of periodic points are thus obtained
$$
v(\langle \gamma_2\rangle)=\{\bm{x}|\ T_2=0,\ T_3-T'_3=0\}
$$
in the case of period 2, and
$$
v(\langle \gamma_3\rangle)=\{\bm{x}|\ T_1=0,\ T_2=0\}
$$
in the case of period 3.

\section{Concluding remarks}

Throughout this paper we have shown many examples of invariant varieties of periodic points by studying rational maps which are known integrable. The problem of finding a criterion which distinguishes an integrable map from nonintegrable one is, however, not easy. As is known, for instance, there are some subtleties involved in defining integrability even when all periodic points are neutral\cite{Berry, Bogomolny}. Nevertheless the existence of invariant varieties of periodic points seems to play an important role in characterizing integrable maps. Before closing this paper we want to discuss relations of our results with those well known in the classical problems of continuous dynamical systems, such as Poincar\'e sections and Painlev\'e properties.\\

It is well known that sections of a periodic orbit of an integrable system draw a curve on a plane of the phase space as we change initial values continuously\cite{Reichl}. Moreover every point on an Arnold torus is neutral in all directions so that orbits are independent from each other. This is strongly related to the fact that an integrable system has sufficiently many constants of motion. When the system is perturbed the periodic points are frozen by Birkhoff's fixed point theorem. The situation is apparently similar to our problem of discrete maps. We must, however, take into account of the difference in the meaning of periodicity in these two cases. The period of the map is the number of steps before it returns, while in the case of the Poincar\'e sections it is the number of sections of an orbit before it hits the same point on the plane. In the map there is no notion of orbit. 

In order to compare these two objects directly we may introduce a parameter $\delta$ which specifies the interval between two steps of the map. If we rescale the dynamical variables $X_j$'s properly we shall obtain various integrable differential equations in the limit of $\delta\rightarrow 0$\cite{Hirota, HTI}. During this standard procedure, however, we encounter the difficulty of distinguishing two points in a periodic map. Consider, for example, a map of period 3
$$
\bm{X}^{(1)}\rightarrow \bm{X}^{(2)}\rightarrow \bm{X}^{(3)}\rightarrow \bm{X}^{(1)}\rightarrow\cdots
$$
After $t=3N\delta$ we will find the point at $\bm{X}^{(1)}$, but see also $\bm{X}^{(2)}$ and $\bm{X}^{(3)}$ at $t+\delta$ and $t+2\delta$, almost the same time in the limit of $\delta\rightarrow 0$.

Instead of introducing $\delta$ we may interpolate the discrete map by certain smooth functions. In doing this we must be very careful not to destroy integrability. If we are considering maps of the Toda family we know how to manage. Namely we write every dynamical variable in terms of a $\tau$ function of the KP hierarchy. If we simply identify the number of steps $n$ with continuous time $t$ the interpolation is established. Assuming our view is correct we should be able to relate our algebraic varieties obtained in this paper with solutions of the KP hierarchy given by hyperelliptic functions. The famous Poincar\'e section of the 3-point Toda lattice found in \cite{Ford}, for example, must be reproduced as a slice of this variety.\\

The Painlev\'e test for an ordinary 2nd order differential equation tells us that it is nonintegrable if there exists a branch point singularity which depends on initial values. The maps we examined in this paper contain differential equations whose integrability is guaranteed by this test in proper continuous limit of $\delta\rightarrow 0$ \cite{Hirota}. Therefore we expect some correspondence to the Painlev\'e test to remain in our approach. But, again, the procedure of taking the limit $\delta\rightarrow 0$ obscures the meaning of the period of the map.\\

Finally let us comment on other works related in some way with our results. Recently Tsuda\cite{Tsuda} discussed the QRT map from a geometrical viewpoint and obtained many interesting results. In particular all periodic maps of QRT are classified corresponding to particular combinations of the parameters. In the paper \cite{HY} the authors derived a number of difference equations whose solutions are limited to periodic functions of arbitrary initial values. These equations, which they called `recurrence equations', include many higher order equations in addition to those derived from the QRT equation. The relations of these works with our results will be interesting to be clarified.

A simpler model which presents the same property that we discussed in this paper was invented by S.Onozawa. Using that model he explored more details of the transition between integrable and nonintegrable maps\cite{Onozawa}. 
\\

\noindent
{\bf Acknowledgement}

The idea of this study arose from discussions with Dr. Katsuhiko Yoshida some years ago. We would like to thank him and also to Mr. Show Onozawa for many interesting discussions. We would like to express our special thanks to Professor Martin Guest who gave us many important suggestions to improve mathematical description as well as English sentences throughout this paper. We also thank Prof. Akira Shudo, Prof. Minoru Omote and Dr. Shigeki Matsutani for valuable comments.

\noindent
{\bf Appendix: 
$\tau$ functions, Toda lattice and Lotka-Volterra maps}\\

Here we explain some relations between $\tau$-functions of the KP theory, the Toda lattice and the Lotka-Volterra maps. All invariants of the series of the Lotka-Volterra map will be presented explicitly.

The $\tau$ function of the KP theory satisfies the Hirota-Miwa equation,
\begin{equation}
\tau_j(l+1,m+1)\tau_j(l,m)-\tau_j(l+1,m)\tau_j(l,m+1)-\tau_{j+1}(l,m)\tau_{j-1}(l+1,m+1)=0,
\label{Hirota-Miwa}
\end{equation}
with $j,l,m\in \bm{Z}$. If we define
\begin{equation}
I_j(l,m):={\tau_{j-1}(l,m)\tau_j(l+1,m)\over\tau_j(l,m)\tau_{j-1}(l+1,m)},\qquad 
V_j(l,m):={\tau_{j+1}(l,m)\tau_{j-1}(l,m+1)\over\tau_j(l,m)\tau_j(l,m+1)}
\label{I, V by tau}
\end{equation}
they satisfy the 2 discrete time Toda lattice equations:
\begin{eqnarray*}
I_{j}(l,m+1)V_j(l+1,m)&=&I_{j+1}(l,m)V_j(l,m),\\
I_j(l,m+1)+V_{j-1}(l+1,m)&=&I_j(l,m)+V_j(l,m).
\end{eqnarray*}
The reduction to one discrete time Toda lattice is achieved by introducing new time variable $t=l+m$ and writing 
$$
I_j(l,m)=I_j^{(t)},\quad V_j(l,m)=V_j^{(t)}.
$$
To obtain a finite system we simply impose conditions $I_{j+N}^{(t)}=I_j^{(t)},\ V_{j+N}^{(t)}=V_j^{(t)}$. We call it the $N$ point Toda map. 

If we define $X_j^{(t)}$ by
\begin{equation}
I_j^{(t)}=(1-X_{2j-1}^{(t)})(1-X_{2j}^{(t)}),\qquad V_j^{(t)}=X_{2j}^{(t)}X_{2j+1}^{(t)}
\end{equation}
they define the Lotka-Volterra map
\begin{equation}
X_j^{(t+1)}(1-X_{j-1}^{(t+1)})=X_j^{(t)}(1-X_{j+1}^{(t)}).
\label{LV equation}
\end{equation}
When the label $j$ of $X_j^{(t)}$ changes from 1 to $d$ and $X_{j+d}^{(t)}=X_j^{(t)}$ is satisfied we call the map a $d$ dimensional Lotka-Volterra map. If $d$ is even, the Lotka-Volterra map is equivalent to the $d/2$ point Toda map.

These are the results known in the literature, especially in \cite{HTI}. Integrability of these maps is obvious since their solutions can be written explicitly in terms of the $\tau$ functions. In \cite{HTI} a simple method is given to find invariants of the $N$ point Toda map. We extend this method, in the following, to the case of the $d$ dimensional Lotka-Volterra map and give invariants explicitly. To this end we first define two matrices
$$
R(t)=\left(\matrix{1-X_1^{(t)}&1&0&\cdots&0\cr
0&1-X_2^{(t)}&1&&\vdots\cr
\vdots&&\ddots&&0\cr
0&\cdots&0&1-X_{d-1}^{(t)}&1\cr
1&0&\cdots&0&1-X_d^{(t)}\cr}\right),
$$$$
L(t)=\left(\matrix{0&0&\cdots&0&X_d^{(t)}&1\cr
1&0&&&0&X_1^{(t)}\cr
X_2^{(t)}&1&0&\cdots&&0\cr
0&X_3^{(t)}&1&0&\cdots&\cr
\vdots&&&\ddots&&\vdots\cr
0&\cdots&0&X_{d-1}^{(t)}&1&0\cr}\right).
$$
A straightforward calculation will show that the Lotka-Volterra equations (\ref{LV equation}) are equivalent to the following matrix formula,
\begin{equation}
L(t+1)R(t+1)=R(t)L(t).
\label{LR=RL}
\end{equation}

To find invariants we define
$$
A(t):=L(t)R(t)=\left(\matrix{1&0&&\cdots&0&p_d&1\cr
1&1&0&&&0&p_1\cr
p_2&1&1&0&\cdots&&0\cr
0&p_3&1&1&\cdots&\cr
\vdots&&&&\ddots&&\vdots\cr
0&&&&1&1&0\cr
0&&\cdots&0&p_{d-1}&1&1\cr}\right),
$$
where we used the notation $p_j:=X_j^{(t)}(1-X_{j-1}^{(t)})$. Since it satisfies
$$
A(t+1)=L(t+1)R(t+1)=R(t)A(t)R^{-1}(t),
$$
eigenvalues of $A$ are invariant. If we write
$$
\det(A-\lambda)=(-1)^{d-1}\sum_{k=0}^dH_{k}(\lambda-1)^k,
$$
the set of coefficients $H_0,H_1,...,H_d$ are also invariant. Comparing the both sides we find
\begin{equation}
H_k=\left\{\begin{array}{cl}
1-(-1)^dp_1p_2\cdots p_d,&\quad k=0\cr
\displaystyle{{\sum}'_{j_1,j_2,...,j_k}p_{j_1}p_{j_2}\cdots p_{j_k},}&\quad k=1,2,...,[d/2]\cr
0,&\quad k=[d/2]+1,...,d-1\cr
-1,&\quad k=d.\cr
\end{array}\right.
\label{H_k}
\end{equation}
Here $[d/2]=d/2$ if $d$ is even and $[d/2]=(d-1)/2$ if $d$ is odd. The prime in the summation $\sum'$ of (\ref{H_k}) means that the summation must be taken over all possible combinations of $j_1,j_2,...,j_k$ but not including direct neighbours. For example, in the case of $d=5$,
$$
H_2=p_1p_3+p_2p_4+p_3p_5+p_4p_1+p_5p_2,
$$
from which we obtain (\ref{H_2 of d=5}). The number of independent invariants of $d$ dimensional Lotka-Volterra map is $[d/2]+1$. Since $H_0$ can be represented by other $H_k$'s and
$$
r=X_1^{(t)}X_2^{(t)}\cdots X_d^{(t)},
$$
it is convenient to use $r$ instead of $H_0$.
\end{document}